\documentclass[twocolumn]{aastex631}

\usepackage[T1]{fontenc}
\usepackage{amsmath}
\usepackage{graphicx}

\shorttitle{CI Tau}
\shortauthors{Shimizu et al.}
\graphicspath{{./}{figures/}}

\begin{document}

\title{High-contrast Imaging around a 2\,Myr-old CI Tau with a Close-in Gas Giant}

\author{Toshinori Shimizu}
\affiliation{Department of Astronomy, Graduate School of Science, The University of Tokyo, 7-3-1 Hongo, Bunkyo-ku, Tokyo 113-0033, Japan}

\author[0000-0002-6879-3030]{Taichi Uyama}
   \affiliation{Infrared Processing and Analysis Center, California Institute of Technology, 1200 E. California Blvd., Pasadena, CA 91125, USA}
    \affiliation{NASA Exoplanet Science Institute, Pasadena, CA 91125, USA}
    \affiliation{National Astronomical Observatory of Japan, 2-21-1 Osawa, Mitaka, Tokyo 181-8588, Japan}

\author[0000-0003-4676-0251]{Yasunori Hori}
\affiliation{National Astronomical Observatory of Japan, 2-21-1 Osawa, Mitaka, Tokyo 181-8588, Japan}
\affiliation{Astrobiology Center, 2-21-1 Osawa, Mitaka, Tokyo 181-8588, Japan}

\author[0000-0002-6510-0681]{Motohide Tamura}
\affiliation{Department of Astronomy, Graduate School of Science, The University of Tokyo, 7-3-1 Hongo, Bunkyo-ku, Tokyo 113-0033, Japan}
\affiliation{National Astronomical Observatory of Japan, 2-21-1 Osawa, Mitaka, Tokyo 181-8588, Japan}
\affiliation{Astrobiology Center, 2-21-1 Osawa, Mitaka, Tokyo 181-8588, Japan}

\author[0000-0003-0354-0187]{Nicole Wallack}
\affiliation{Earth and Planets Laboratory, Carnegie Institution for Science, Washington, DC 20015}
\affiliation{Division of Geological \& Planetary Sciences, California Institute of Technology, MC 150-21, Pasadena, CA 91125, USA}




\begin{abstract}

Giant planets around young stars serve as a clue to unveiling their formation history and orbital evolution.
CI Tau is a 2\,Myr-old classical T-Tauri star hosting an eccentric hot Jupiter, CI Tau\,b. The standard formation scenario of a hot Jupiter predicts that planets formed further out and migrated inward. A high eccentricity of CI Tau b may be suggestive of high-$e$ migration due to secular gravitational perturbations by an outer companion. Also, ALMA 1.3\,mm-continuum observations show that CI Tau has at least three annular gaps in which unseen planets may exist.
We present high-contrast imaging around CI Tau taken from Keck/NIRC2 $L^{\prime}$-band filter and vortex coronagraph that allows us to search for an outer companion. We did not detect any outer companion around CI Tau from angular differential imaging (ADI) using two deep imaging data sets. The detection limits from ADI-reduced images rule out the existence of an outer companion beyond $\sim30$\,au that can cause the Kozai-Lidov migration of CI Tau\,b. Our results suggest that CI Tau\,b may have experienced Type II migration from $\lesssim 2$\,au in Myrs.
We also confirm that no planets with $\geq 2-4\,M_\mathrm{Jup}$ are hidden in two outer gaps.

\end{abstract}



\section{Introduction}
\label{Introduction}
    Planets around young stars provide a clue to understanding planetary orbital evolution before the dispersal of the protoplanetary disk. Recently, some close-in giant planets (hereafter hot Jupiters) around young stellar objects (YSOs) whose ages are $\lesssim 10$\,Myr have been reported \citep[e.g., V830 Tau\,b,][]{2016Natur.534..662D}. The standard formation scenario of a hot Jupiter is the core accretion model \citep[e.g.][]{1996Icar..124...62P} which predicts that planets formed farther away from a star, followed by inward orbital migration. There are several mechanisms that cause planetary migration. A gap-opening planet migrates inward by planet-disk interactions, also referred to as type II migration \citep[e.g.][]{1986ApJ...309..846L}. Since the disk gas can damp the eccentricity of a planet, a migrating planet has a low eccentricity of $\lesssim 0.1$ \citep{2015ApJ...812...94D,2017MNRAS.464L.114R, 2018MNRAS.474.4460R}. Dynamical processes driven by secular perturbations of outer companions, such as the Kozai-Lidov mechanism \citep[e.g.][]{2007ApJ...669.1298F,2011Natur.473..187N} and gravitational scattering among planets \citep[e.g.][]{1996Natur.384..619W}, excite the eccentricity of an inner planet. The orbit of an eccentric planet decays and becomes circular due to tidal interactions.
    In general, the formation of a close-in gas giant through ``high-e migration'' occurs in $\sim 0.1-1$\,Gyr after disk dispersal.
    
    CI Tau is a 2\,Myr-old classical T-Tauri star, which is known as one of the YSOs \citep{age2014}. We note that the age of this system has uncertainties  of $\sim \pm\,0.5$\,Myr because of its young age. CI Tau has an eccentric hot Jupiter, CI Tau b with a predicted eccentricity of $e=0.28 \pm 0.16$ \citep{Johns-Krull2016}.
    A high eccentricity of a hot Jupiter around YSOs with ages of 1--10\,Myr is a challenge to the two migration scenarios. 
    ALMA continuum observations also indicate that the CI Tau disk has at least three annular gaps at $\sim$13, 39 and 100 au in 1.3\,mm \citep{Clarke2018}. These gaps could indicate the existence of additional unseen outer companions. In addition to this, ALMA band 7 (0.89\,mm) observations revealed that there is a gap at $\sim50$\,au in the intensity map of $^{13}$CO\,$J = 3-2$ emission, but this gap structure is unlikely to be due to a reduction of the gas surface density induced by an unseen planet \citep{2021MNRAS.501.3427R}. The near infrared polarimetric image of VLT/SPHERE did not detect any gap near 50\,au in the scattered light because of a large vertical scale height. \citep{2022A&A...658A.137G}.
    
    Kozai-Lidov migration of a planet could explain the eccentricity of young CI Tau\,b.
    The existence of a massive outer companion near $\sim 10-30$\,au may trigger the Kozai-Lidov mechanism in $\sim 1-10$\,Myr.  
    A planet perturbed by an outer companion, however, undergoes eccentricity damping due to gas drag in the residual disk.
    These two competing processes obscure the origin of CI Tau\,b.
    
    How massive outer companions are around CI Tau allows us to constrain the orbital evolution of CI Tau\,b. In this study, we performed high-contrast imaging with Keck/NIRC2 and conducted angular differential imaging \citep[ADI;][]{2006ApJ...641..556M} to explore the existence of an outer companion. Even non-detection of outer companions can rule out the existence of massive companions above the detection limits of direct imaging surveys. Section \ref{Observations} describes our observations. Section \ref{Data reduction} describes data  analysis and section \ref{Result} shows the results. Section \ref{Discussion} describes discussions about planetary evolution. Finally, Section \ref{Summary} summarizes this research.

\section{Observations}
\label{Observations}
    We used two archival data of CI Tau taken from Keck/NIRC2 with the $L^{\prime}$ band on October 21 (Obs.A) and December 23 (Obs.B), 2018, which were part of a protoplanetary disk survey (Wallack et al. {\it in prep}). The data were obtained with the vector vortex coronagraph \citep{Mawet_2017} and we adopted an inner working angle (IWA) = 125 mas \citep[i.e., 12.5 pixels][]{IWA1_2017,IWA2_2017}. 
    The total exposure time of Obs.A and B were 3960\,$s$ and 2610\,$s$, which consist of 132 and 87 frames, respectively.
    The total angular rotations of Obs.A and B were $\sim161^{\circ}$ and $\sim182^{\circ}$, respectively. Table \ref{table1} summarizes the details of both observations.
   
    \begin{table*}
        \caption{Observation Log}
            \begin{tabular*}{17.8cm}{@{\extracolsep{\fill}}ccccccc}
            \hline \hline
            Date &Single Exposure Time [s]& Coadds & Frames & Total Exposure Time [s] & Band&PA rotation\\\hline
            2018 Oct. 21 & 0.75 & 40&132 & 3960 &$\mathrm{L^{\prime}}$& $161^{\circ}$\\
            2018 Dec. 23 & 0.75 & 40 &87 & 2610 &$\mathrm{L^{\prime}}$& $182^{\circ}$\\ \hline
            \end{tabular*}
        \label{table1}
    \end{table*}
    
    We also used unsaturated frames as a PSF reference of the host star. These frames of both data were comprised of 100 coadded 0.008\,$s$-frames. We determined that the FWHM of the host star PSF observed on October 21 and December 23 was 8.7 pix and 8.9 pix (pixel scale is 9.971 mas/pix), respectively, from two-dimensional Gaussian fitting. The photometry of the host star was also determined, which was used in later analyses.

\section{Data reduction}
\label{Data reduction}
    We used ADI reduction to search for an outer companion with the pyKLIP package \citep{pyKLIP2015}, which is a python library to remove the stellar halo and speckles by making a reference PSF from the data sets. This package implements the Karhunen-Loeve Image Processing algorithm (KLIP; \cite{KLIP2012}) and KLIP-FM \citep{KLIPFM2016}. The pre-processing pipeline that we used \citep{Xuan2018} includes a flat field correction, removing bad pixels, sky-subtraction, and position-registration. First, we made an averaged image from all the frames, which were subtracted from the science frames. The purpose of this process is to remove the halo of the host star and make it easier to detect planets. We then applied ADI reduction to the residual frames while searching for the best pyKLIP parameters.
    By injecting two fake sources at $r = 30\,\mathrm{pixel}$ at an angle of $\sim0^{\circ}$ and $\sim180^{\circ}$, we determined the best parameters that could retrieve the fake sources at the best signal-to-noise (SN) ratios. Details about the noise calculations are explained in Section \ref{Result}.
    Parameters ``KL'' and ``movement'' define the number of basis vectors and the minimum amount of movement (in pixel) of an astrophysical source used for a reference PSF image, respectively.
    The ``annuli'' parameter sets the number of annuli and we divided the FoV into annular regions where post-processing was applied separately. The width of the annulus is (OWA - IWA)/'annuli', where OWA corresponds to the outer working angle; OWA of these images = 2.9\,arcsec (292 pix).
    
    \begin{table}
        \caption{the parameters of pyKLIP}
        \begin{tabular*}{8.3cm}{@{\extracolsep{\fill}}cccccc}
        \hline \hline
        & IWA [pix] & KL & movement [pix] & annuli \\\hline
        Oct. 21 & 12.5 & 20 & 0.3 & 9  \\ 
        Dec. 23 & 12.5 & 15 & 0.5 & 8  \\ \hline
        \end{tabular*}
        \label{table2}
    \end{table}
    
\section{Result}
\label{Result}
\begin{figure*}
        \begin{tabular}{cc}

        \begin{minipage}{0.45\hsize}
        \includegraphics[scale=0.65]{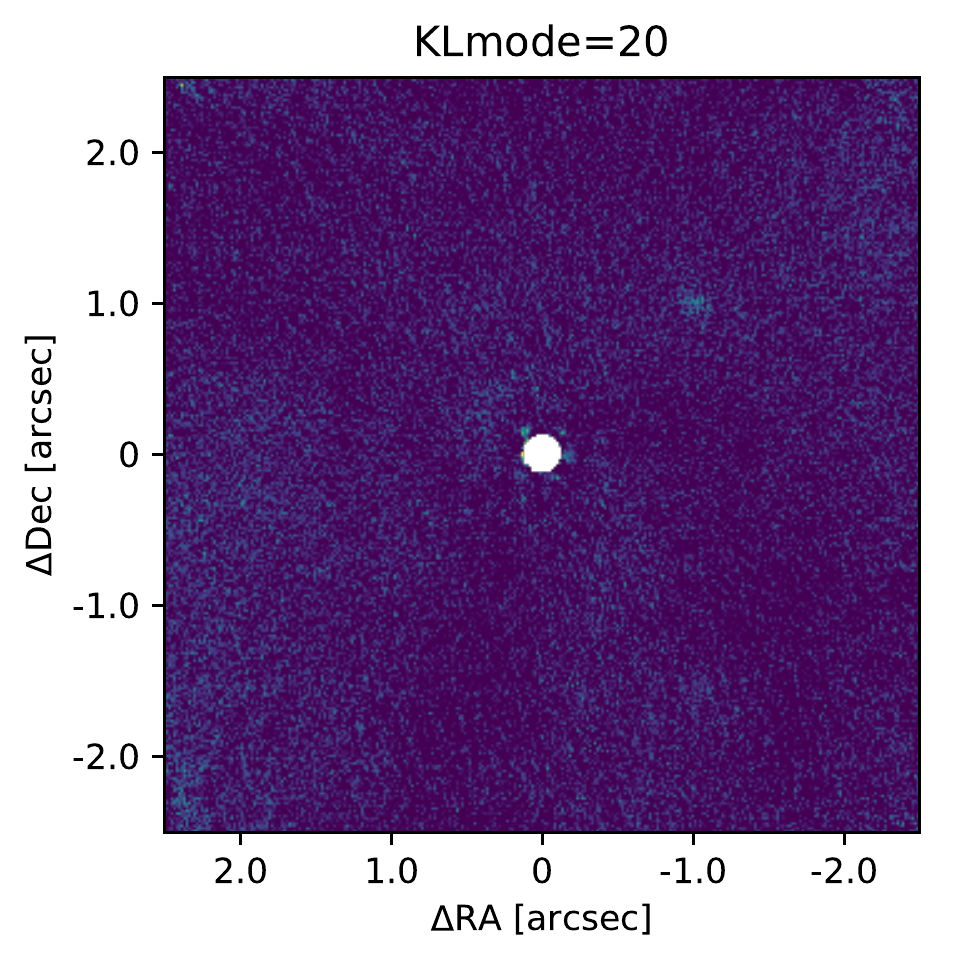}
        \end{minipage}

        \begin{minipage}{0.45\hsize}
        \includegraphics[scale=0.65]{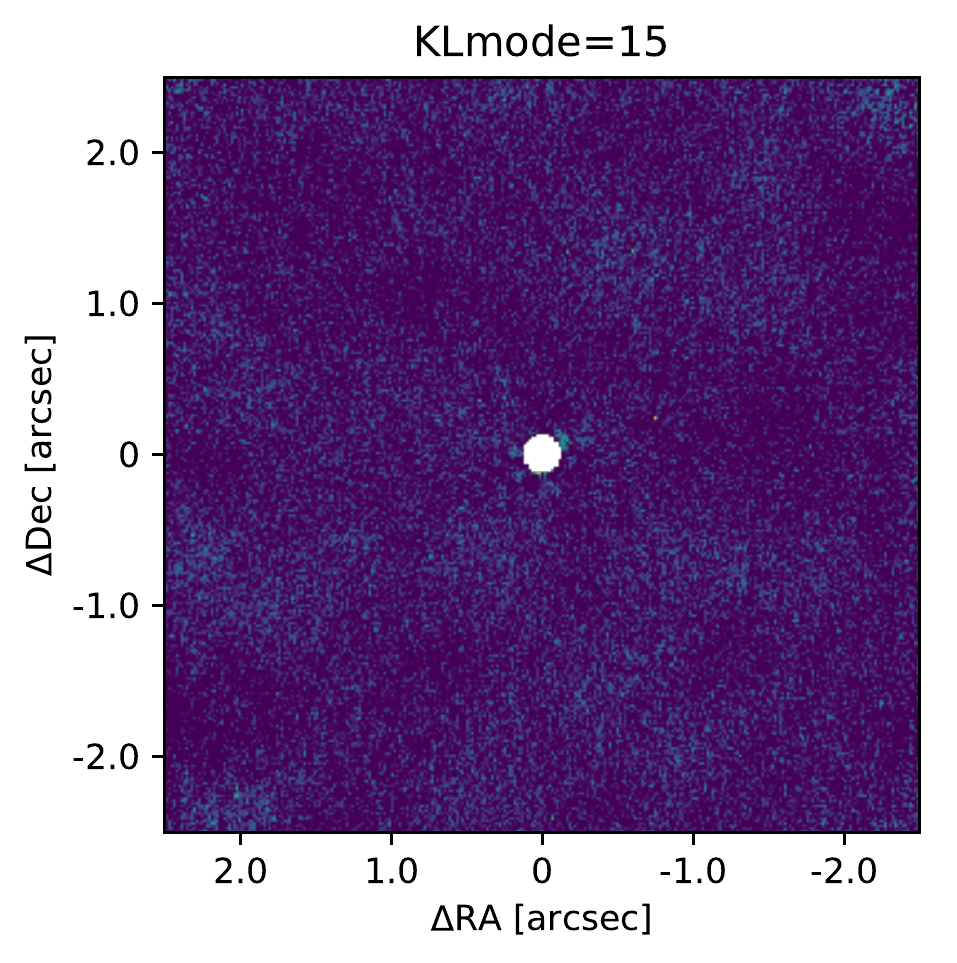}
        \end{minipage}

        \end{tabular}
        \caption{ADI-reduced images of CI Tau. Left is Oct. 21, 2018 (KL = 20) and right is Dec. 23, 2018 (KL = 15). The white area in the center is the coronagraph masking. North is up, and east is to the left. The October data are suffered from an artificial emission within optics (seen at ($\sim$ -1.0", +1.0")). See text for more details.}
        \label{fig1}
    \end{figure*}

\begin{figure*}
        \begin{tabular}{cc}

        \begin{minipage}{0.45\hsize}
        \includegraphics[scale=0.65]{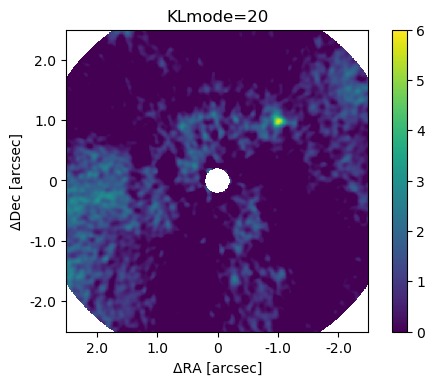}
        \end{minipage}

        \begin{minipage}{0.45\hsize}
        \includegraphics[scale=0.65]{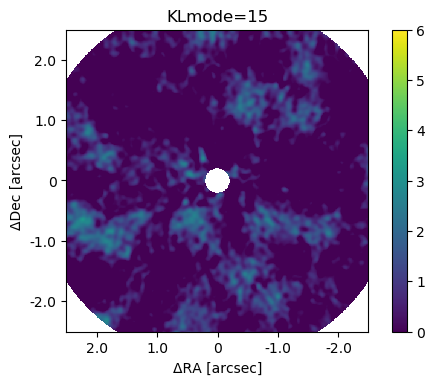}
        \end{minipage}

        \end{tabular}
        \caption{SN maps calculated by the images after ADI reduction (see Figure \ref{fig1}). North is up, and east is to the left.}
        \label{fig2}
    \end{figure*}

    Figure \ref{fig1} shows the ADI-reduced results (KL = 20 and 15, respectively). Using the ADI-reduced images, we created the SN map to determine whether or not an outer companion was detected (Figure \ref{fig2}).
    We convolved the image with a 2D-Gaussian kernel whose $\sigma$ is FWHM/2.35. After this, we calculated the standard deviation within FHWM/2-width annular regions, which provides a noise radial profile after the ADI reduction. From this noise profile, we created a noise map and a SN map by dividing the ADI result by the noise map.
    We found a bright point-like source (SN\,$>5$) at Northwest in the left panel of Figs. 1 and 2.
    Compared to the star's PSF, the varying brightness and the extended shape made it difficult to remove through the ADI reduction. We identified this point-like source as dust inside the instrument from the pre-processed frames.
    We did not detect any other companion candidates in Figures \ref{fig1} and \ref{fig2}.

    
    
\begin{figure}[h]
        \includegraphics[scale=0.44]{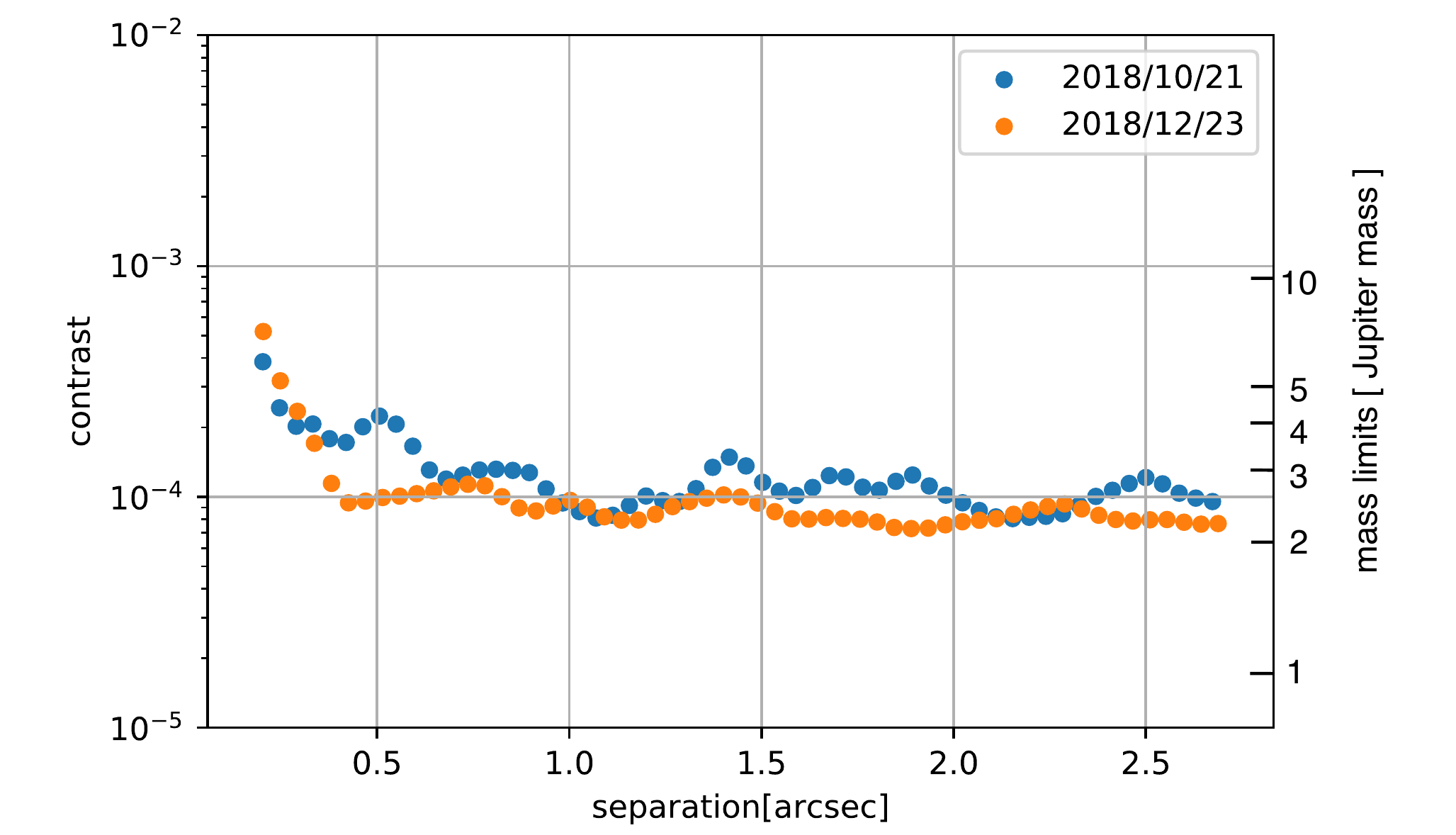}
        \caption{5-$\sigma$ contrast and mass detection limits derived from the ADI-reduced results of each observation. The mass detection limits is converted from the contrast limits using CONDO03 model \citep{CONDO2003}. The age and distance of CI Tau were assumed to be 2\,Myr
        and 160\,pc \citep{2020yCat.1350....0G}}
        \label{fig3}
    \end{figure}

   Although we did not detect a companion, we can still utilize our observation to  get a constraint on the mass of a potential outer companion.
   For this purpose, we performed an injection test to investigate self-subtraction that represents the degree of signal loss due to the ADI reduction. And then, we derived the 5$\sigma$-detection limits of a planet-star contrast as a function of distance from the host star by using the pyKLIP tool. 
   This tool used a Gaussian low pass filter to smooth out high frequency noise; we adopted was FWHM/2.35 for $\sigma$ of a Gaussian. After that, we calculated standard deviation in each annular region to get a noise radial profile and we derived contrast by comparing this to the peak value of the reference PSF.
   We converted the contrast limits to the detection limits of a planet using the $L^{\prime}$ flux of CI Tau (derived from interpolating {\it Spitzer} photometries \citep{2005PASP..117..978R}) and CONDO03 model \citep{CONDO2003}, assuming that the age of CI Tau was 2 Myr. Figure \ref{fig3} shows the 5$\sigma$-detection contrast and mass limits. The contrast limit of $10^{-4}$ at $0\farcs5$ corresponds to $\sim2 M_{\mathrm{Jup}}$ at $r > 50 \mathrm{au}$.
   The weather conditions at the time of the October observation were below average. 
   As a result, we achieved a better detection limit from the December data than the October data, although the October data provided the longer total exposure time.
    
\section{Discussion}
\label{Discussion}
    \subsection{The existence of a potential outer companion}\label{sec:dis2}

    We discuss the existence of an outer companion around CI Tau, using the detection limits of Keck/NIRC2 imaging. 
    The period of eccentricity oscillation ($P_{\mathrm{precess}}$) by the Kozai-Lidov mechanism \citep{Holman1997} is given by 
    \begin{align}
        \label{1}
        P_{\mathrm{precess}} \simeq P_{\mathrm{orb}} \left(\frac{M_\mathrm{s} + M_\mathrm{p}}{M_\mathrm{b}}\right) \left(\frac{a_\mathrm{b}}{a_\mathrm{p}}\right)^{3} (1-e_\mathrm{b})^{3/2},
    \end{align}
    where the subscripts, s, p, and b represent the host star, planet, and outer companion, respectively, 
    $P_{\mathrm{orb}}$ is the orbital period of the outer companion, $M$ is the mass, $a$ is the semi-major axis, and $e$ is the eccentricity. Note that $M_\mathrm{s} > M_\mathrm{b} > M_\mathrm{p}$ and we assumed $e_{\mathrm{b}}=0$ to obtain the maximum of $P_{\mathrm{precess}}$.
    We obtain the minimum mass of an outer companion that causes Kozai-Lidov oscillations of the planet's eccentricity from $ t_{\mathrm{age}} > P_{\mathrm{precess}}$:
    \begin{align}
        \label{2}
        M_\mathrm{b} > \frac{P_{\mathrm{orb}}}{t_{\mathrm{age}}} \left( \frac{a_\mathrm{b}}{a_\mathrm{p}} \right)^3 M_\mathrm{s},
    \end{align}
    where $t_{\mathrm{age}}$ is the age of the host star.
    
    The apsidal precession of a planet due to general relativity (GR) effects takes place.
    The eccentricity excitation of a planet is suppressed due to GR effects as well as tidal interactions between a planet and a host star. The timescale of GR precessions \citep[see e.g.][]{2013ApJ...773..187N} is estimated as
    \begin{align}
    \label{3}
        P_{\mathrm{GR}} &= \frac{2\pi c^2(1-e^2_\mathrm{p})a_{\mathrm{p}}^{5/2}}{3(GM_{\mathrm{s}})^{3/2}} \notag\\
        & = 33.6 \left(\frac{P_{\mathrm{orb}}}{1\,\mathrm{yr}}\right) \left(\frac{a_{\mathrm{p}}}{1\,\mathrm{au}}\right) \left(\frac{M_\mathrm{s}}{M_\odot}\right)^{-1}\,\,\,\mathrm{Myr},
    \end{align}
    where $c$ is the speed of light, $G$ is the gravitational constant, and $M_{\odot}$ is the solar mass.
    The orbit of a planet can become eccentric by the Kozai-Lidov mechanism if $P_\mathrm{precess} < P_\mathrm{GR}$, namely,
    \begin{align}
    \label{4}
        M_{\mathrm{b}} > \frac{P_{\mathrm{orb}}}{P_{\mathrm{GR}}} \left( \frac{a_{\mathrm{b}}} {{a_{\mathrm{p}}}} \right)^3 M_{\mathrm{s}}.
    \end{align}

\begin{figure}[h]
        \includegraphics[scale=0.6]{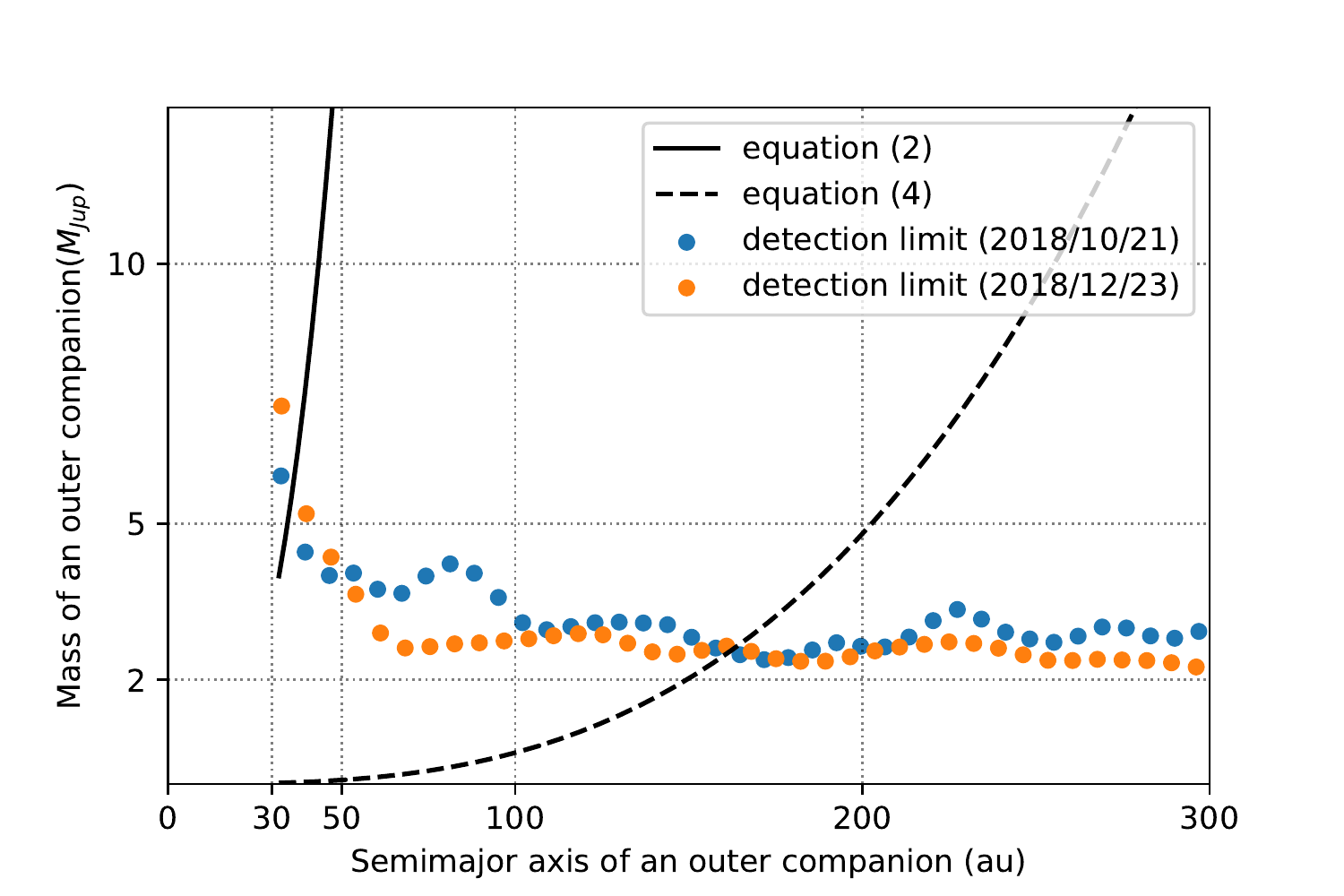}
        \caption{Detection limits of an outer companion in Keck/NIRC2 $L^{\prime}$-band images. Solid and dashed lines show the minimum mass of the outer companion that satisfies equations (\ref{2}) and (\ref{4}), respectively.}
        \label{fig4}
    \end{figure}

     CI Tau\,b has the mass of $11-12\,M_\mathrm{Jup}$ ($M_\mathrm{p}\sin{i} = 8.08\pm1.53\,M_\mathrm{Jup}$) and the orbital period of $P_\mathrm{orb}=8.9891\, \mathrm{days}$ \citep{Johns-Krull2016}.
     We consider that CI Tau\,b initially formed near the snow line ($\sim 2.4\ \mathrm{au}$) and then, moved to the current location.
    Figure \ref{fig4} shows the detection limits of Keck/NIRC2 $L^{\prime}$-band imaging and the minimum mass of the outer companion that satisfies equations (\ref{2}) and (\ref{4}).
    The detection limits of high-contrast imaging are lower than the minimum mass of an outer companion beyond $\sim 30$\,au for the Kozai-Lidov mechanism. We confirmed that there is no outer companion outside $\sim 30$\,au that caused the Kozai-Lidov migration of CI Tau\,b. 
    We, however, cannot rule out the possibility that an outer companion within $\sim 30$\,au caused the Kozai-Lidov mechanism.


    Three annular gaps in the CI Tau disk detected by the ALMA sub-mm observations suggested the existence of planets at $\sim$13, 39 and 100\,au \citep{Clarke2018}.
    The detection limits of our direct imaging confirmed that there is no outer companion with $\geq 2\,M_\mathrm{Jup}$ at 100\,au and with $\geq 4\,M_\mathrm{Jup}$ at 39\,au. In other words, smaller planets with $< 2-4\,M_\mathrm{Jup}$ are likely to open the two gaps at 39\,au and 100\,au.
    The annular gap at 13\,au is located inside the IWA of Keck/NIRC2. We need deep high-contrast imaging with a smaller IWA in order to peer into the innermost gap.
    Nevertheless, planets in the gaps should be dynamically stable because orbital separations between gaps are larger than $2\sqrt{3}r_\mathrm{Hill}$ \citep{1993Icar..106..247G}, where $r_\mathrm{Hill}$ is the mutual Hill radius of neighboring planets.
    
 \subsection{The origin of an eccentricity of CI Tau\,b}
    We discuss the orbital evolution of a hot Jupiter CI Tau\,b.
    As mentioned in Section \ref{sec:dis2}, an outer companion may exist within $\sim 30$\,au.
    Such an outer companion, however, should have strongly affected the orbit of a planet hidden in the inner gap.
    The Kozai-Lidov oscillations of CI Tau\,b may be incompatible with the locations of inner gaps around CI Tau.  
    Here, we revisit type II migration scenario of an eccentric CI Tau\,b.
    
    A migrating planet suffers from the eccentricity damping due to gas drag.
    The migration rate of a planet ($\tau_{\mathrm{a}} \equiv a_\mathrm{p}/u_\mathrm{p}$) is related to the viscous drift of disk gas.
    Assuming $\alpha$-viscosity model with $\alpha=10^{-3}$, we obtain the viscous drift timescale ($\tau_{\mathrm{\nu}} \equiv a_\mathrm{p}/u_\mathrm{vis}$):
    \begin{align}
        \tau_\mathrm{\nu}=\frac{a_\mathrm{p}}{u_\mathrm{vis}}&=\frac{2}{3}\left(\frac{a_\mathrm{p}}{h}\right)^2\alpha^{-1}\Omega_\mathrm{K}^{-1} \notag\\
    \label{8}
        & = 2.5\times10^{4}\left(\frac{a_\mathrm{p}}{1\mathrm{au}}\right)^{3.25}\left(\frac{M_\mathrm{s}}{M_\odot}\right)^{-\frac{1}{2}},
    \end{align}
    where $u_\mathrm{vis}$ ($= 3\nu/2a_\mathrm{p}$, where $\nu$ is the kinetic viscosity) is the radial velocity of viscous gas accretion,
    $h$ is the density scale height in disk $z$-direction and $\Omega_\mathrm{K}$ is the Keplerian angular velocity. We adopted the CI Tau's disk model \citep{CITau2021}.
    According to \cite{Kanagawa2018}, the ratio of $\tau_{\mathrm{\nu}}$ to $\tau_{\mathrm{a}}$ is written as
    \begin{align}
      \frac{\tau_\mathrm{\nu}}{\tau_\mathrm{a}} =&\frac{u_\mathrm{p}}{u_\mathrm{vis}} = 100\frac{h_\mathrm{p}}{a_\mathrm{p}}\frac{\Sigma_\mathrm{p} a_\mathrm{p}^2}{M_\mathrm{p}} \notag\\
      =&\,2.2\times10^{-1}\left(\frac{a_\mathrm{p}}{1\mathrm{au}}\right)^{0.125}\left(\frac{M_\mathrm{p}}{M_\mathrm{J}}\right)^{-1} \notag\\
    \label{9}
      &\exp\left[-\left(\frac{a_\mathrm{p}}{0.08\mathrm{au}}\right)^{-2}\right]\exp\left[-\left(\frac{a_\mathrm{p}}{2.5\mathrm{au}}\right)^{2}\right],
    \end{align}
    where $u_\mathrm{p}$ is the migration speed of a planet and
    $h_\mathrm{p}$ and $\Sigma_\mathrm{p}$ are the disk scale height and the surface density of disk gas at the location of a planet. 
    The two exponential terms in equation (\ref{9}) represent a disk cutoff \citep[see][]{CITau2021}.

    We calculated $\tau_\mathrm{a}$ from equations (\ref{8}) and (\ref{9}), assuming that CI Tau\,b started to migrate from the snow line. 
    The age of CI Tau is estimated to be $\sim 2$ Myr.
    As the surface density of disk gas decreases with time, we consider that the CI Tau's disk had 10 times as high as $\Sigma_\mathrm{p}$ given in \citet{CITau2021}.
    The migration timescale of a planet at 1.8\,au (1.6\,au for the CI Tau's disk with a cutoff) is comparable to the age of CI Tau. 
    CI Tau\,b may have formed near 1.8\,au because the snow line was located in the inner region around a T Tauri star\citep[e.g.][]{2011Icar..212..416M,2021ApJ...916...72M}.
    Type II migration still could explain the current orbit of CI Tau\,b.
    The eccentricity damping of a migrating planet, however, occurs more quickly than the orbital decay \citep[e.g.][]{1980ApJ...241..425G}. Therefore, a high eccentricity of CI Tau\,b may not be compatible with type II migration scenario unless the eccentricity of CI Tau\,b may be overestimated.
    Neither the Koza-Lidov mechanism nor the orbital instability that triggers planet-planet scattering is likely to occur. If CI Tau\,b were in an eccentric orbit, our results would be a unique testbed to revisit the eccentricity change of a Type II migrating planet in a disk.
    Further follow-up observations are needed to understand the origin of this elusive hot Jupiter with the next generation of ground-based telescopes which have a larger aperture and a smaller IWA coronagraph.

\section{Summary}
\label{Summary}
We have presented high-contrast imaging with Keck/NIRC2 $L^{\prime}$-band and vortex coronagraph around CI Tau to search for outer companion. We conducted ADI reduction using the pyKLIP package to remove the stellar halo and speckles. We did not detect any outer companion from either of the two observations.
The $5\sigma$-contrast detection limits confirmed that there is no an outer companion outside $\sim 30$\,au that causes the Kozai-Lidov mechanism. Although three annular gaps at $\sim$13, 39 and 100\,au in the CI Tau disk were detected by the ALMA observations, the innermost annular gap at 13\,au is located inside the IWA. For two outer gaps, we confirmed that there is no outer companion with $\geq 2\,M_\mathrm{Jup}$ at 100\,au and with $\geq 4\,M_\mathrm{Jup}$ at 39\,au.
Note that we cannot rule out the possibility that an outer companion exists within $\sim 30$\,au.
Our results suggest that CI Tau\,b formed near a snowline and experienced Type II migration.
This scenario, however, may not be compatible with a high eccentricity of CI Tau\,b because the eccentricity of a migrating planet is damped more quickly than the orbital decay.
Therefore, further follow-up observations are required to determine the eccentricity of CI Tau\,b more accurately and observe the inner region with deep high-contrast imaging that allows us to search for a smaller companion, if any.

\begin{acknowledgments}
We are grateful to Dimitri Mawet and Garreth Ruane for sharing the NIRC2 data taken in their observational programs.
T.U. is supported by Grant-in-Aid for Japan Society for the Promotion of Science (JSPS) Fellows and JSPS KAKENHI Grant No. JP21J01220.
YH was partly supported by JSPS KAKENHI Grant Number 18H05439.
M.T. is supported by JSPS KAKENHI grant Nos.18H05442,15H02063,and 22000005.
This research is partially supported by NASA ROSES XRP, award 80NSSC19K0294.

The data presented herein were obtained at the W. M. Keck Observatory, which is operated as a scientific partnership among the California Institute of Technology, the University of California and the National Aeronautics and Space Administration. The Observatory was made possible by the generous financial support of the W. M. Keck Foundation.
The authors wish to recognize and acknowledge the very significant cultural role and reverence that the summit of Maunakea has always had within the indigenous Hawaiian community.  We are most fortunate to have the opportunity to conduct observations from this mountain.
\end{acknowledgments}

\bibliographystyle{aasjournal}
\bibliography{adslib}    



\end{document}